\documentstyle[prl,psfig,preprint,aps]{revtex}
\begin{document}

\draft
\tighten
\title{Generalized seniority from random Hamiltonians }

\author{C.~W.~Johnson$^a$, G.~F.~Bertsch$^{b,c}$, D.~J.~Dean$^d$, and
I.~Talmi$^{b,e}$ }
\address{$^a$Department of Physics and Astronomy,
Louisiana State University, Baton Rouge Louisiana 70803\\
$^b$Institute for Nuclear Theory,
University of Washington, Seattle, WA 98195\\
$^c$ Department of Physics, University of Washington, Seattle, WA 98195\\
$^d$Physics Division, Oak Ridge National Laboratory, Oak Ridge, Tennessee
37831,\\
and Department of Physics and Astronomy, University of Tennessee, Knoxville,
Tennessee 37996\\
$^e$Weizmann Institute of Science, Rehovot, Israel}

\maketitle

\begin{abstract}

We investigate the generic pairing properties of shell-model
many-body Hamiltonians
drawn from ensembles of random two-body matrix elements. Many
features of pairing that are commonly attributed to the
interaction are in fact seen in a large part of the ensemble
space.  Not only do the spectra show evidence of pairing with
favored $J=0$ ground states and an energy gap, but the relationship
between ground state wave functions of neighboring nuclei show signatures
of pairing
as well.  Matrix elements of pair
creation/annihilation operators between ground states tend to be
strongly enhanced.  Furthermore, the same or similar pair operators connect
several ground states along an isotopic chain. This
algebraic structure is reminiscent of the generalized seniority model.
Thus pairing may be encoded to a certain extent in the Fock space
connectivity of
the interacting shell model even without specific features of the
interaction required.

\end{abstract}


\section{Introduction}

Pairing in fermion systems is a ubiquitous phenomenon that appears
among fermions as diverse as electrons, nucleons, and $^3$He atoms.
In the nuclear shell model this is usually explained as due to
strong, attractive matrix elements of the two-body effective interaction
between $J=0$ pair states.
In the Fermi liquid model it is explained as a consequence
of the interaction being attractive at certain momentum and
energy transfers.  In this paper we wish to demonstrate that features of
pairing arise from a very large ensemble of two-body interactions and,
hence, are independent, to a large extent, from the specific character
of the interaction.  Pairing
may be favored simply as a consequence of the two-body nature of the
interaction and the way it connects the Fock space wave functions of
the noninteracting Fermion system.  We examine this possibility by
numerical studies of the many-body system governed by a two-body
Hamiltonian taken from a random ensemble.  We consider nucleons in
a spherical shell-model space for which $J=0$ states have a special
significance.  We thus take ensembles of two-body Hamiltonians
that respect angular momentum, but otherwise are as general as possible.

In an earlier paper\cite{jo98}, we examined spectral features of pairing with
one such ensemble and found that two signatures were present in the
preponderance of Hamiltonians:  the ground state tends to have $J=0$
angular momentum, and there tends to be a gap between that state
and the higher states in the spectrum.  Another important feature shown
in [1] was referred to as phonon collectivity.  Ground states with
$J=0$ had, on the average, rather large matrix elements of a
single-nucleon operator with the first excited $J=2$ states.  This is a
spectral characteristic of pairing, but it may occur in other situations
as well. Why pairing should be favored in a general ensemble is not
clear.   It has been suggested that time-reversal invariance is
responsible, but that symmetry has been found to be unnecessary
\cite{bi99}. In any case, energy spectra are only part of the properties
affected by
pairing. The term ``pairing'' implies that the ground state
can be approximated or modeled by a
 condensate of pairs of fermions coupled, in the interacting
shell model, to zero angular momentum; that is,
the ground state is approximately of the form
$(\sum_{\alpha,m} c_\alpha a^\dagger_{\alpha, m}
a^\dagger_{\alpha, -m})^N|0\rangle$. Furthermore, under some
conditions the matrix elements of pairing operators have an
algebraic structure, as described
by the seniority model and its generalizations (see \cite{talmi}).
Along these lines, we consider here two
signatures
that explicitly probe the wave functions:

$\bullet$ a strong pair-transfer amplitude;
that is, there is a large matrix element of the pair annihilation operator
between the ground states of the nuclei with $A-2$ and $A$ nucleons;

$\bullet$ considering an
isotopic chain of nuclides together, $A$, $A-2$, $A-4$, etc., the seniority
model and its generalizations predict that the same pair annihilation
operator that takes one from the $A$ ground state to the $A-2$
ground state will also take one from the
$A-2$ to the $A-4$ ground state, etc.

We also consider here an additional energy signature:

$\bullet$ in an isotopic chain containing both even and odd $A$, the
even isotopes have systematically greater binding energies.

Besides considering these three additional signatures of pairing,
we will examine some other ensembles to see how robust our results
are.  The calculations are performed in
two shell model spaces:  the ``$sd$" space consisting of the orbitals
with angular momentum $j=1/2, 3/2, 5/2$ which can accommodate up to
12 identical particles, and the ``$pf$" space which has in addition the
$j=7/2$ shell and accommodates up to 20 identical particles.  We use
the Glasgow-Los Alamos\cite{haxton}
 and ANTOINE shell-model codes\cite{caurier}
to calculate the many-body
wave functions and observables in these shell-model spaces.  The
Hamiltonian is specified by the single-particle and two-particle
matrix elements.  Except for one ensemble discussed below, we
set the single-particle matrix elements to zero.

\section{Two-Body Random Interactions}

For the two-body matrix elements, we choose a basis of two-body
states, labeled by
$\alpha$, which has good angular momentum $J$.
There are 63
independent two-body matrix elements in the $sd$ space and 195 in
the $pf$ space, including both neutrons and protons.  We define
an ensemble of two-particle Hamiltonians requiring that the
ensemble be invariant under changes in the basis of two-particle
states.  This can be achieved by taking the matrix elements to be
Gaussian distributed about zero with the widths possibly depending
on $J$,

\begin{equation}
\langle V_{\alpha,\alpha'}^2\rangle
= c_{J_\alpha} ( 1+ \delta_{\alpha,\alpha'} )  \bar{v}^2.
\end{equation}
$$
\langle V_{\alpha,\alpha'}V_{\beta,\beta'}\rangle =
0, \,\,\,(\alpha,\alpha')\ne(\beta,\beta')
$$
Here $\bar{v}$ is an overall energy scale that we generally ignore (except
for scaling single-particle energies for the RQE-SPE defined below).
The coefficients $c_{J}$ then define the ensemble.
We emphasize that $J$ refer to quantum
numbers of  {\it two-body} states, and not of the final
many-body states (typically 4-10 particles).

We now discuss the choices of ensembles, which may be specified by the
$c_{J}$ coefficients
and the single-particle Hamiltonian, if present.  In our earlier work we
employed the `RQE' ensemble defined below, but it is important to examine
other ensembles to see how robust the results are.
The ensembles are:

1. {\bf RQE} (Random Quasiparticle Interaction). Here $c_{J} = (2J+1)^{-1}$.
The relation between the $c_{J}$ came from imposing an additional
invariance on the ensemble, that it be the same for the particle-particle
interaction as for the particle-hole interaction\cite{jo98}.  The
RQE gives a larger variance to $J=0$ matrix elements than to the others.
Note that even though the variance is larger, the matrix elements
are both attractive and repulsive, so there is no bias toward pairing
by the traditional mechanism of an attractive two-particle interaction.

2. {\bf TBRE} (Two-body Random Ensemble).  Here $c_{J} =$ constant.
Historically, this was the
first two-particle random ensemble to be employed in studying statistical
properties of many-particle
spectra \cite{fr70}.

3. {\bf RQE-NP} (Random Quasiparicle Ensemble-No Pairing).  This
is the same as the RQE ensemble, except all $J=0$ two-body matrix
elements are set equal to zero.  This ensemble will show clearly
whether the $J=0$ channel matrix elements are needed at all to produce
the signatures of pairing.  (It is known \cite{talmi}, albeit not
widely appreciated, that it is possible to  have interactions that
are diagonal in seniority without any explicit
$J=0$ pairing interaction.)

4. {\bf RQE-SPE} (Random Quasiparticle Ensemble-with Single-Particle
Energies). All the previous ensembles had the single-particle
energies set to zero. Realistic interactions do have nonzero single-particle
energies, and these can, in principle, affect pairing properties, at the
very least by creating large shell gaps.  For calculations in the {\it
sd} shell we take here single-particle energies from the Wildenthal
interaction
\cite{wildenthal}, scaling
$\bar{v}=3.84$ MeV
so as to best match the widths of the two-particle matrix elements.
For the {\it pf} shell we use single particle energies from the
modified KB3 interaction \cite{kb3} and scaled $\bar{v}=4.43$ MeV.

\section{Results}

For the specific calculations,
we considered  4, 6, and 8 neutrons in the $sd$ space; we label these
as the corresponding shell-model systems, $^{20}$O, $^{22}$O,
and $^{24}$O respectively;
we caution the reader that this labeling can be misleading as we have
deliberately put in as little physics of those systems as possible.
We also considered
4, 6, 8, and 10 neutrons in the $pf$ space: $^{44}$Ca, $^{46}$Ca,
$^{48}$Ca, and $^{50}$Ca.
Finally, we included systems with nontrivial
isospin, considering in the $sd$-shell 4 protons and 4, 6, and 8 neutrons:
$^{24,26,28}$Mg, respectively. In these cases $c_{J} \rightarrow c_{JT}$.
Thus for the RQE, $c_{JT}=(2J+1)^{-1}(2T+1)^{-1}$.
For each of these systems, and for each of the
ensembles described above, we computed at least 1000 samples.

{\bf Spectral signatures}.  Table 1 presents
the fraction of each ensemble that yields a
$J=0$ ground state for the above systems.  For purposes of comparison,
the fraction of the
total many-body states that are $J=0$ and $J=2$ states is also given.
If the ground state spins reflected only the size of the $J$ subspace,
there would be more $J=2$ than $J=0$, contrary to our findings.

In addition to a predominance of $J=0$ ground states, such states
are pushed down relative to the rest of the spectrum. An example
is shown in Fig.\ 1.  Note that for spectrum 1(a), the
$J=0$ ground state is separated from the excited states by an amount
large compared to the average level spacing, while for the
case of a $J > 0$ ground state,  1(b),  the separation of the ground
state and the average level spacing is similar.

This is shown in
more detail in Fig.\ 2.  Here we define $s$ to be the spacing between
the ground state and the first excited state, scaled by the local
level spacing $D$, defined as the ensemble-averaged
spacing between
the first and second excited states.  Because these states in general
do not have the same quantum numbers such as total $J$, one would
expect the level spacing to be described by a Poisson distribution,
where the probability of finding a spacing $s$ is given by
$P(s) = \exp(-s/D)$ \cite{brody}.
For cases where the ground state $J \neq 0$,
the Poisson distribution describes the distribution of $s$ extremely
well, as one would predict.   For those cases where the ground state
$J=0$, however, the distribution is much broader. It is somewhat
approximated by a Poisson, but with $D$ 3 times larger.
We show two cases in Fig.\ 2. The other nuclides and
ensembles yield nearly identical figures. Table II tabulates the average
$\langle s \rangle$ for the various ensembles.  For all cases, a
$J=0$ ground state is pushed down an {\it average} factor of 2.3-3.7
relative to the local level spacing, whereas a $J > 0$ ground state
is, within statistics, not pushed down at all.
Similar results hold for the Mg ensembles.

The third spectral feature is the well-known even-odd staggering of
ground state energies.  Figures 3 illustrates the real world situation
 with the experimental
neutron removal energies $S_n(A) = -E(A)+E(A-1)$ of calcium isotopes
in the range A=45-50.  The larger removal energy of the even isotopes
is associated with their greater binding energy.  We look for
evidence of this in our ensemble spectra of the $pf$ isotope chains
$A=4-10$ as follows.  We first
examined the even members of the chain, requiring that all ground states
have $J=0$.  This is satisfied for
$\sim42\%$ of the members of the RQE ensemble; this is a
much larger than expected value $(0.70)^4=0.25$
that one would obtain from Table 1 assuming that the $J=0$ occurrences
are uncorrelated.  In the generalized seniority model \cite{talmi},
the even-member ground state energies have a quadratic dependence on
$A$, \cite{talmi} eq. (23.20):
\begin{equation}
E_{gs}(A) = a+b A+c A^2.
\label{quadratic_fit}
\end{equation}
We next make a least-squares fit of the selected even-$A$ chains to
this formula.
This is, of course, a fit of 3 parameters to 4 data points and the
description is good.  Examples of the deviations about this fit are plotted in
Fig.\ 4.  We then computed the binding energies for $n=5,7$. The
deviations from (\ref{quadratic_fit}) for $^{47,48}$Ca are plotted in Fig.\ 4,
scaled
to the local level spacing:  $^{48}$Ca exemplifies all the even-$n$ cases,
which are all very similar,
while  $^{45}$Ca yields a plot nearly identical to that of $^{47}$Ca.
Notice that not only are the odd-particle systems
consistently higher in energy, they are pushed up on average by
3 times the local level spacing -- which is entirely consistent with the
results shown in Fig.\ 2 and Table II.  Figure 4 also contains results from
the RQE-NP ensembles in the $pf$ shell.  Even with all $J=0$ matrix
elements set to zero, we find qualitatively similar results.  The effects
are not as dramatic in this case; from Table I one would expect all four
isotopes $^{44,46,48,50}$Ca to have $J=0$ ground states $6.5\%$ of
the time, but in fact this occurs $8.4\pm0.8\%$ of the time.

{\bf Pair-transfer collectivity.}
The spectral and energetic characteristics discussed above are not
the only signatures of pairing; matrix elements of pairing operators
are also very important.  In order to test the hypothesis
that the ground states of these random Hamiltonians can be approximated
by pair condensates, let's follow the example of generalized
seniority and consider the general  pair-annihilation operator
 $S = \sum_j \alpha_j S_j$, where
 $S_j = \sum_{m > 0} (-)^ma_{jm} a_{j-m}$
is  the pair-annihilation operator for the $j$-shell.
Given $S$, the pair-transfer amplitude from the ground state with
$A$ particles to $A-2$ particles is
$\langle A-2 | S | A \rangle$.  One way to probe the wave function is
the {\it pair-transfer fractional collectivity} (defined in
analogy with the phonon fractional collectivity of Ref. \cite{jo98}):
\begin{equation}
f_p = {  \langle A-2 | S | A \rangle^2 \over \langle A |S^\dagger S | A
\rangle}.
\end{equation}
If the states of the system are condensates of the $S^\dagger$ pairs,
then one expects $f_p =1$.

How does one determine the $\alpha_j$?  Because the ensembles are defined
to be invariant on changes of basis, there cannot be a globally
preferred $\alpha_j$.
In principle, we could determine individual $\alpha_j$ for each
ensemble member by maximizing $f_p$ from Eq.\ (3).
However, the variational
condition is rather complicated, and we found satisfactory
evidence of pairing collectivity with a much simpler ansatz.  In analogy to
phonon fractional collectivity used in \cite{jo98},
we set
\begin{equation}
\alpha_j = \langle A-2 | S_j | A \rangle.
\end{equation}

Figure 5 presents the distribution of the $f_p$ for various `nuclides'
and interaction ensembles. The ensemble denoted `GOE' refers to
using two different RQE interactions for the $A$ and $A-2$ wave functions;
one would expect a minimal correlation between their wave functions and indeed
the distribution of $f_p$ is heavily weighted towards zero  for all
nuclides.
For the cases
using the same interaction for the $A$ and $A-2$ wave functions, however,
we get distinctly different results:  a  weighting towards $f_p=1$,
implying an enhanced correlation indicative of a pairing-like condensate.
All our nuclides and ensembles yield similar plots. The results are
summarized in Table III in the form of the average fractional pairing.
Keep in mind that the distributions for GOE have a negative slope,
while for all other ensembles the slope of the distribution is positive.
(We also tabulate, for comparison, the exact $f_p$ for realistic interactions:
the Wildenthal
interaction \cite{wildenthal} in the $sd$ shell and modified
KB3\cite{kb3} in the $pf$ shell.) Thus,
in the cases of Ca and O,
for all these ensembles--even those with the $J=0$ pairing
matrix elements explicitly removed--we see an increased enhanced number
of states with a condensate-like ground state.
The Mg nuclei lie in between the GOE and ensembles of identical
nucleons.
This indicates that the proton-neutron interaction dampens the pairing
collectivity
present in all-neutron systems such as the $^{20-24}$Ca and
$^{44-50}$Ca isotopes.  The difference is likely due to the $T=0$
interaction.

For interactions that are truly diagonal in generalized seniority,
one expects the {\it same} condensate to prevail for $A=2, 4, 6, 8, ...$
valence nucleons \cite{talmi}.  In the language developed above,
let $\alpha_{j}(A)$ be the coefficients computed from $A$ and $A-2$.
The $\{ \alpha_j(A)\}$ can be thought of as vectors, and from
generalized seniority we expect the vectors $\vec{\alpha}(A)$ and
$\vec{\alpha}(A+2)$ to be aligned. To test this idea,
define the scalar product $\vec{\alpha}(A)\cdot \vec{\alpha}(A')
= \sum_j \alpha_j(A) \alpha_j(A')$, where the states and matrix elements
are calculated with the same two-body Hamiltonian.  (One could have
different weightings
or metrics for this scalar product, such as $\sqrt{2j+1}$ or
$1/\sqrt{2j+1}$, but such differences in definition do not change
our results.)    Then plot the distribution of
\begin{equation}
\cos \theta ={ | \vec{\alpha}(A)\cdot \vec{\alpha}(A+2) |
\over \left | \alpha(A) \right |
 \left | \alpha(A+2) \right | }
\end{equation}
Calculations for the O and Ca isotopes using
realistic interactions \cite{wildenthal,kb3}
typically give 0.99 for this correlation
factor, except at shell closures of the $d_{5/2}$ in the
$sd$ shell and $f_{7/2}$ in the $pf$ shell, where it is 0.4-0.5.
For the Mg isotopes with the Wildenthal interaction,
 this factor is $0.7$, indicating that
the likelihood for the same correlated pair to be transferred  along
the chain is somewhat less than the all-neutron case.
If all $T=0$ matrix elements are set to zero, then one recovers
the factor 0.99 for the correlation.
The results for the GOE and RQE ensembles for O and Ca
 are plotted in Fig.\ 6.
For the GOE case we find
a flat distribution -- the pair-transfer amplitudes are uncorrelated, exactly
as one would expect.
However, for the ensembles of random two-body interactions, we
find for the O and Ca chains
a sharp peak at 1, indicating a strong correlation.
The chain $^{28}$Mg$\rightarrow$$^{26}$Mg$\rightarrow$$^{24}$Mg,
plotted in Fig.\ 7b, also
shows a peak at 1, which is a factor of 4 higher than the average bin
height. In contrast, the O and Ca peaks are at least a
factor of 10 above the average
bin height. Thus, the pair transferred in the O and Ca chains is
much more likely to be of the same condensates than the
pair transferred in the Mg case.
The other ensembles yield plots similar to that shown for the RQE.  Curiously
enough, for the RQE-NP ensemble (not shown),
we also find a sharp peak at $\cos \theta=0$,
as well as at $\cos \theta = 1$.

The analysis described in the previous paragraph only considered
`nearest-neighbor' transitions.  If, however, we have an approximate
generalized seniority, then we expect the pair-transfer amplitude vector
to be similar for a whole chain of isotopes.  We compare, for the RQE,
the correlation for the pair-transfer amplitudes starting from $^{50}$Ca
$\rightarrow ^{48}$Ca and computing the correlation, not only with
$^{48}$Ca $\rightarrow ^{46}$Ca, but also with
$^{46}$Ca $\rightarrow ^{44}$Ca.  This correlation shows an enhancement
at the value that is similar to the results shown in Fig.\ 6.  Thus we have
strong
evidence that the pairing condensate is not an arbitrary and local feature,
but persists along an isobaric chain.

\section{Conclusions}

We have considered several random ensembles of two-body Hamiltonians
in the framework of the shell model.  By examining the statistical
properties of the low-lying spectra, as well as pair-transfer
amplitudes, we find pairing behavior occurs frequently
in our ensembles of  two-body interactions.  Thus, pairing is a
robust feature of two-body Hamiltonians.  There seems to be a large class
of two-body interactions leading to pairing which is much wider than
the attractive interactions usually considered.

Besides pairing, there are other features of nuclear spectra that
often occur and can give rise to algebraic structures.  The most
prominent example, rotational bands, is not favored at all by
random Hamiltonians.  Since a rotational band implies an internal
rigidity of the system, this shows that in some sense the random
ensembles describe only Fermi liquid behavior.  In the spherical
shell model, it has been shown that it is the $T=0$ part of the
nuclear effective interaction acting between neutrons and protons
which gives rise to collective spectra like rotational ones.  The
effective $T=0$ interaction has a rather strong quadrupole component
which breaks the seniority coupling scheme.

It is interesting to speculate on the more complex algebraic structures
that have been found in nuclear spectroscopy.  The phenomenologically
successful interacting boson model is based on collective pair transfer
operators in both $J=0$ and $J=2$ (quadrupole) angular momenta.
Since we see no indication of
quadrupole collectivity in the random Hamiltonians, one would have to
introduce from the start some physical features of the interaction.  It might
be that the important physical features could be described very
simply, say, by an attractive surface delta interaction.  One would
then look for the rich variety of observed dynamical symmetries by
adding to the physical component a component from one of the random
ensembles.

\acknowledgements
CWJ and IT thank the Institute for Nuclear Theory and the program on
Algebraic Methods in Many-Body Physics, where much of this work was
initiated and the first drafts of this paper were written.  We had
many stimulating discussions, including A. Dieperink, J. Ginocchio,
F. Iachello, and S. Tomsovic.
This work is supported by
Department of Energy grant numbers  DE-FG02-96ER40985,
DE-FG-06-90ER40561, and
DE-FG02-96ER40963.
Oak Ridge National Laboratory is managed by Lockheed Martin Energy
Research Corp. for the U.S. Department of Energy under contract number
DE-AC05-96OR22464.

\begin{table}
\caption
{Percentage of ground states for selected random ensembles that
have $J=0$ for our target nuclides, as compared to
the percentage of  all states in the model spaces that have
these quantum numbers. (Statistical error is approximately
$1$--$3\%$.)  Entries with dashes ``--'' were not computed.
}
\begin{tabular}{|c|c|c|c|c|c| c|}
 Nucleus & RQE & RQE-NP & TBRE &  RQE-SPE    & $J=0$  &  $J=2$ \\
           &   & & &  &   (total space)  &    (total space) \\
\hline
  $^{20}$O    & $68\%$ & $50\%$ & $50\%$  & $49\%$  & $11.1\%$ &
$14.8\%$  \\
  $^{22}$O    & $72\%$ & $68\%$ & $71\%$ &  $77\%$  &  $9.8\%$ &
$13.4\%$ \\
   $^{24}$O   & $66\%$ & $51\%$ & $55\%$  & $78\%$  & $11.1\%$ &
$14.8\%$ \\
   \hline
 $^{44}$Ca    & $70\%$ & $46\%$ &   $41\%$ & $70\%$   & $5.0\%$  &
$9.6\%$ \\
 $^{46}$Ca    & $76\%$ & $59\%$ & $56\%$  & $74\%$    & $3.5\%$  &
$8.1\%$ \\
  $^{48}$Ca   & $72\%$ & $53\%$ &  $58\%$  & $71\%$    & $2.9\%$  &
$7.6\%$  \\
 $^{50}$Ca    & $65\%$ & $45\%$ & $51\%$  & $61\%$ & $2.7\%$  &
$7.1\%$  \\
\hline
$^{24}$Mg     & $66\%$ &  --    &  $44\%$ & $54\%$ & $4\%$   &
 $16\%$     \\
$^{26}$Mg     & $62\%$ & $52\%$ &  $48\%$ & $56\%$ & $4\%$   &
 $15\%$     \\
$^{28}$Mg     & $59\%$ & $46\%$ &  $44\%$ & $54\%$ & $4\%$   &
 $16\%$     \\
\end{tabular}
\end{table}

\begin{table}
\caption
{Average gap between $J=0$ ground state and first excited
states, $\langle s \rangle$, scaled by the local level
spacing (computed from the 1st and 2nd excited states).
The same quantity computed for $J>0$ ground states is
between 0.9 and 1.2 for all cases considered.
}
\begin{tabular}{|c|c|c|c|c|}
 Nucleus & RQE & RQE-NP & TBRE &  RQE-SPE    \\
\hline
  $^{20}$O    & 2.7 & 2.5 &  2.3 &  2.3  \\
  $^{22}$O    & 3.2 & 2.8 &  2.8 &  3.4  \\
   $^{24}$O   & 2.9 & 2.5 &  2.3 &  3.7  \\
   \hline
 $^{44}$Ca    & 3.1 & 2.6 & 2.4 &  3.1 \\
 $^{46}$Ca    & 3.8 & 3.2 & 3.0 &  3.6 \\
  $^{48}$Ca   & 3.4 & 3.0 & 3.5 &  3.5 \\
 $^{50}$Ca    & 3.5 & 3.0 & 3.0 &  3.4
\end{tabular}
\end{table}

\begin{table}
\caption
{Average value of fractional pair-transfer collectivity,
$f_{\rm pair}$, between nuclides $A$ and $A-2$.  `Realistic'
= Wildenthal interaction for $sd$ shell nuclides and
KB3 interaction for $pf$ shell nuclides.  GOE denotes
pair-transfer amplitudes between random wave functions;
that is, $A$ and $A-2$ were computed using different
members of the RQE ensemble.}
\begin{tabular}{|c|c|c|c|c|c|c| }
 Nucleus & Realistic & GOE & RQE & RQE-NP & TBRE &  RQE-SPE    \\
 initial$\rightarrow$ final &     &     &        &      &            \\
\hline
  $^{24}$O $\rightarrow$ $^{22}$O
  & 0.99     & 0.25 & 0.77 & 0.75  & 0.78  & 0.86   \\
  $^{22}$O $\rightarrow$ $^{20}$O
  & 0.86     & 0.22 & 0.65 & 0.59  & 0.62  & 0.77   \\
   \hline
  $^{50}$Ca $\rightarrow$ $^{48}$Ca
  & 0.98     & 0.032 & 0.57  &  0.42  & 0.47 & 0.58   \\
  $^{48}$Ca $\rightarrow$ $^{46}$Ca
  & 0.86     & 0.036  & 0.51 & 0.34  & 0.38 & 0.53    \\
  $^{46}$Ca $\rightarrow$ $^{44}$Ca
  & 0.94     & 0.070 & 0.48 & 0.28  & 0.30 & 0.48   \\
\hline
  $^{28}$Mg $\rightarrow$ $^{26}$Mg
   & $0.57$  &           &  $0.26$   & $0.15$  &  &  $0.27$  \\
  $^{26}$Mg $\rightarrow$ $^{24}$Mg
   & $0.72$  &           &  $0.39$   & $0.27$  &  &  $0.47$
\end{tabular}
\end{table}

\bibliographystyle{try}

\begin{figure}
\label{fig1}
  \begin{center}
    \leavevmode
    \parbox{0.7\textwidth}
      {\psfig{file=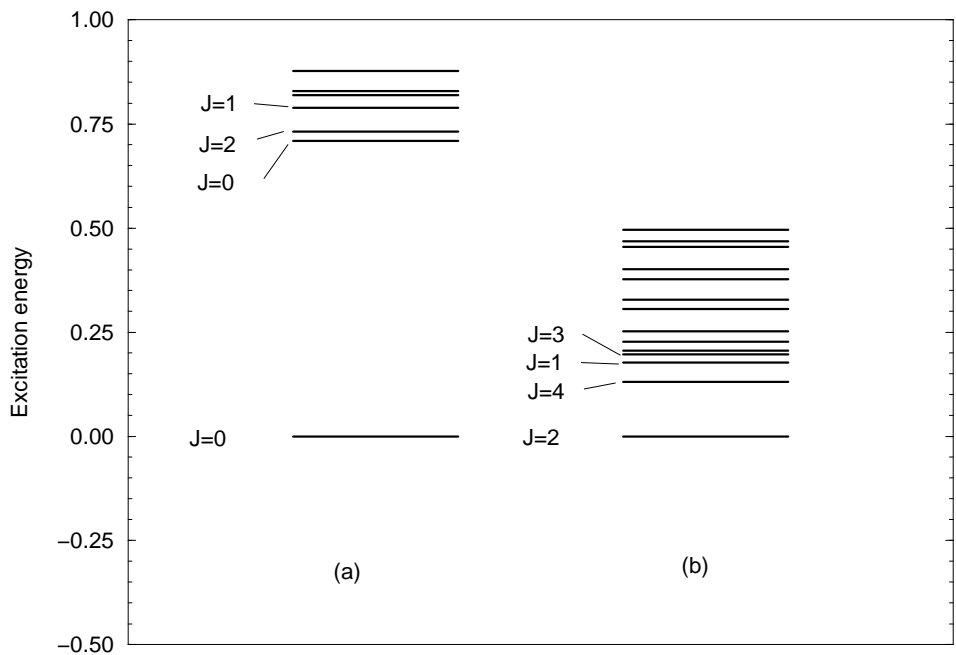,width=0.9\textwidth,angle=270}}
    \end{center}
\caption{ `Typical' spectra for
$^{46}$Ca  with an RQE Hamiltonian: a) an example having $J=0$
ground state; b) an example with $J\neq 0$ ground state.  Note the
absence of a ground-state gap in the $J\neq 0$ case. }
\end{figure}
\newpage

\begin{figure}
\label{fig2}
  \begin{center}
    \leavevmode
    \parbox{0.7\textwidth}
      {\psfig{file=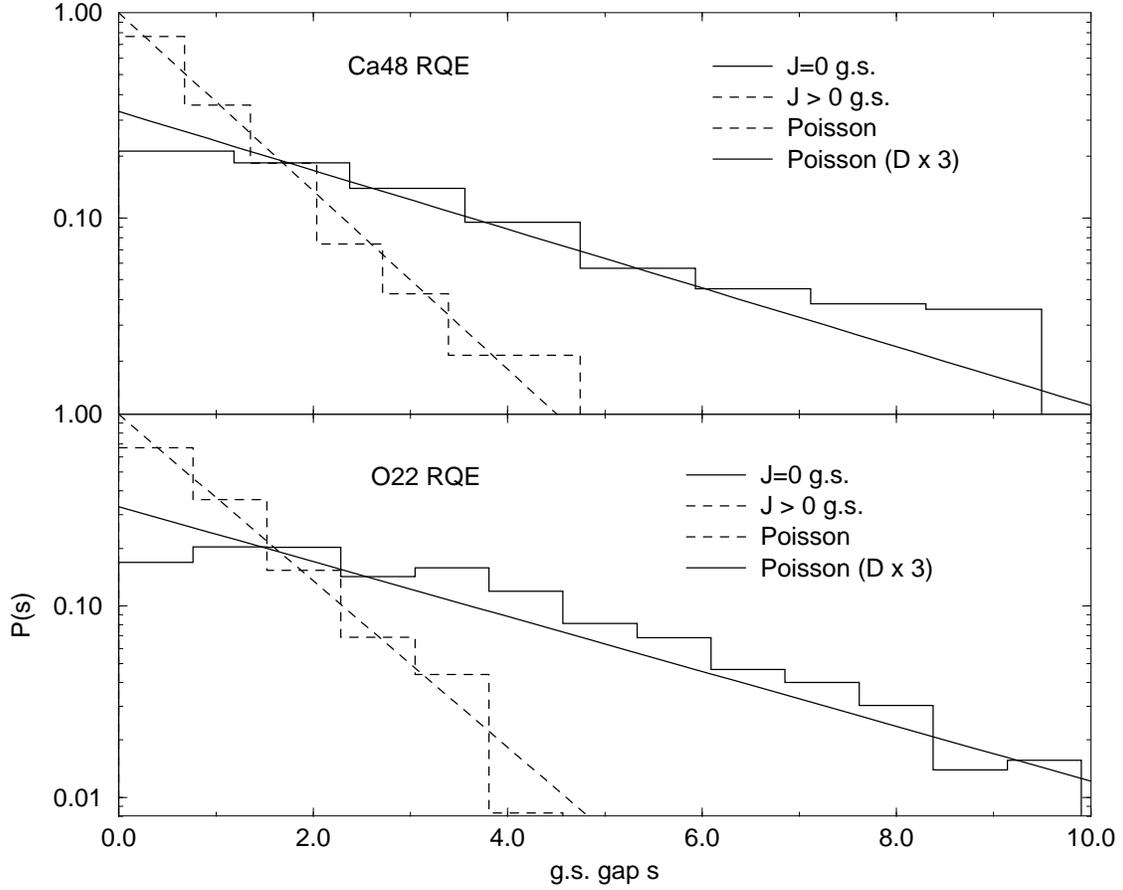,width=0.9\textwidth,angle=270}}
    \end{center}

\caption{ Distribution of ground state gaps, $s = E_1 -E_0$,
in the spectrum of $^{46}$Ca.  Energies are scaled to the
average
local level density defined as the inverse level spacing between
the first and second excited states, $D= \langle E_2 -E_1\rangle$
(averaged over the ensemble).
Dashed
histogram is the distribution for the cases in which the
ground state has nonzero spin, and the dashed curve is the
expected Poisson distribution.  The solid histogram shows the
case for $J=0$ in the ground state.  This is also rather
well fitted by a Poisson curve (solid), but in this case with
an average level spacing 3 times larger.
}
\end{figure}

\begin{figure}
\label{fig3}

  \begin{center}
    \leavevmode
    \parbox{0.9\textwidth}
      {\psfig{file=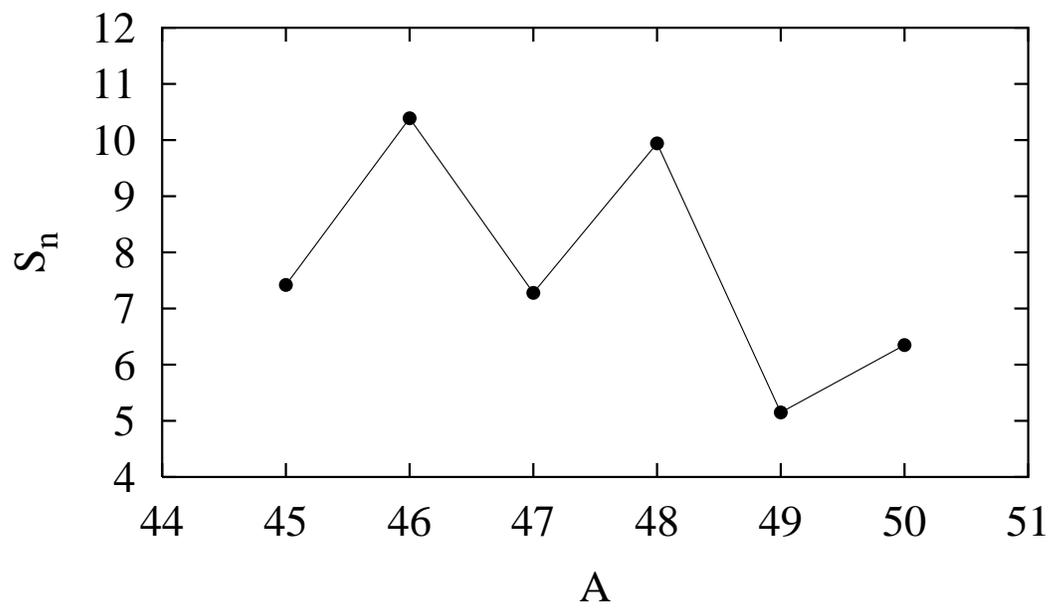,width=0.9\textwidth}}
    \end{center}
\caption{Experimental neutron separation energies of Ca isotopes in
the range A=45-50.  }
\end{figure}
\newpage
\begin{figure}
\caption{Even-odd staggering effect in the RQE and RQE-NP for
4--10 neutrons in the $pf$-shell.  }
\label{fig4}
  \begin{center}
    \leavevmode
    \parbox{0.9\textwidth}
      {\psfig{file=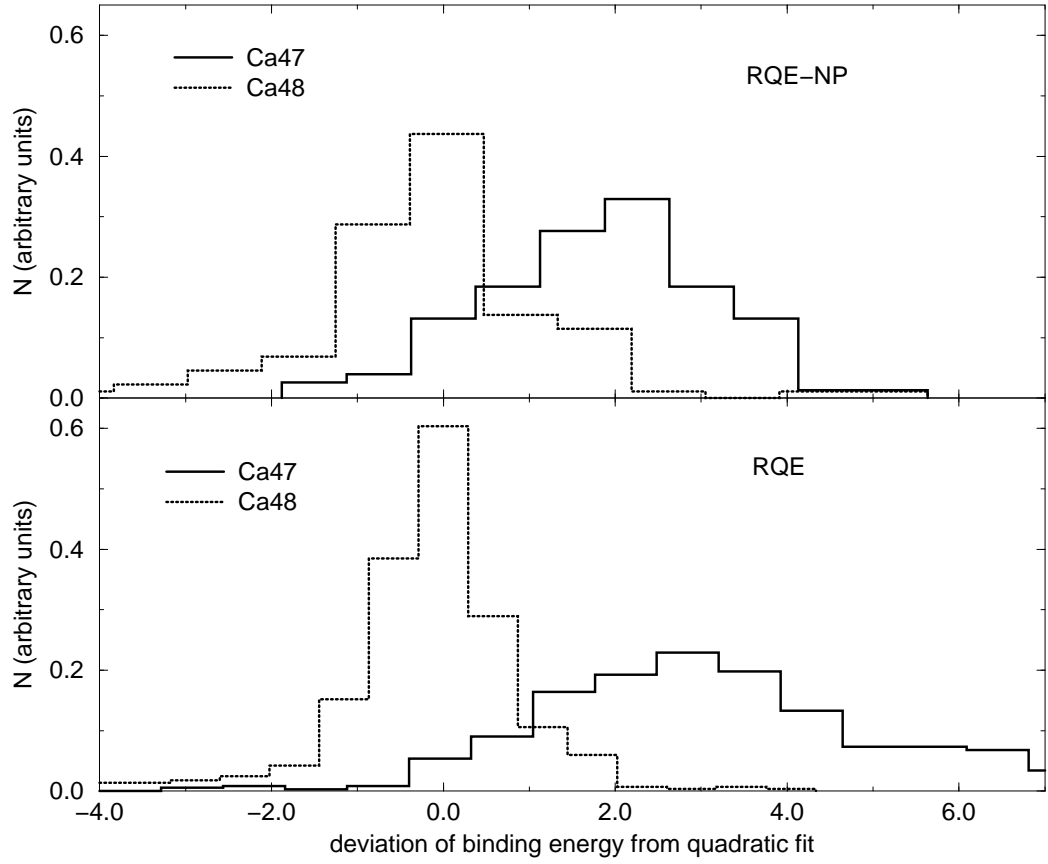,width=0.9\textwidth,angle=270}}
    \end{center}
\end{figure}
\newpage
\begin{figure}
\caption{Distribution of fractional pair-transfer
collectivity, $f_p$, for
selected isotopes and ensembles.}
  \begin{center}
    \leavevmode
    \parbox{0.7\textwidth}
      {\psfig{file=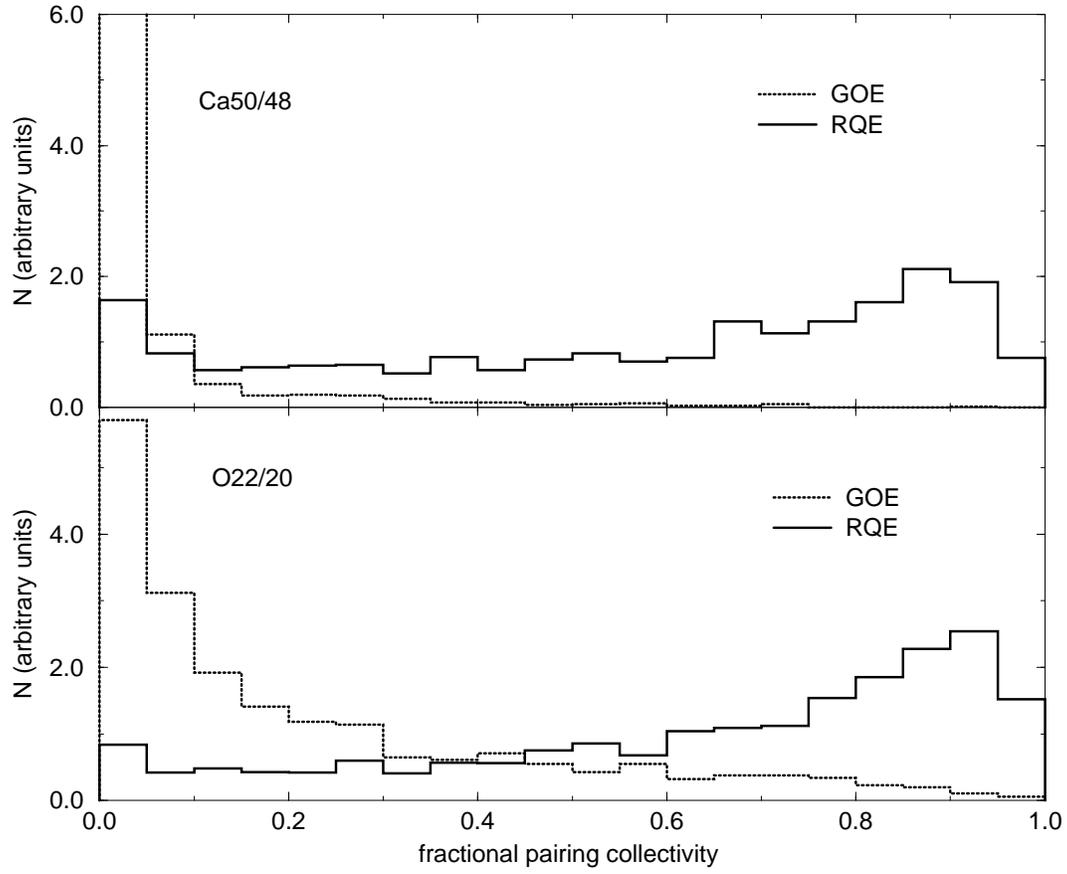,width=0.9\textwidth,angle=270}}
    \end{center}
\end{figure}

\begin{figure}
  \begin{center}
    \leavevmode
    \parbox{0.7\textwidth}
      {\psfig{file=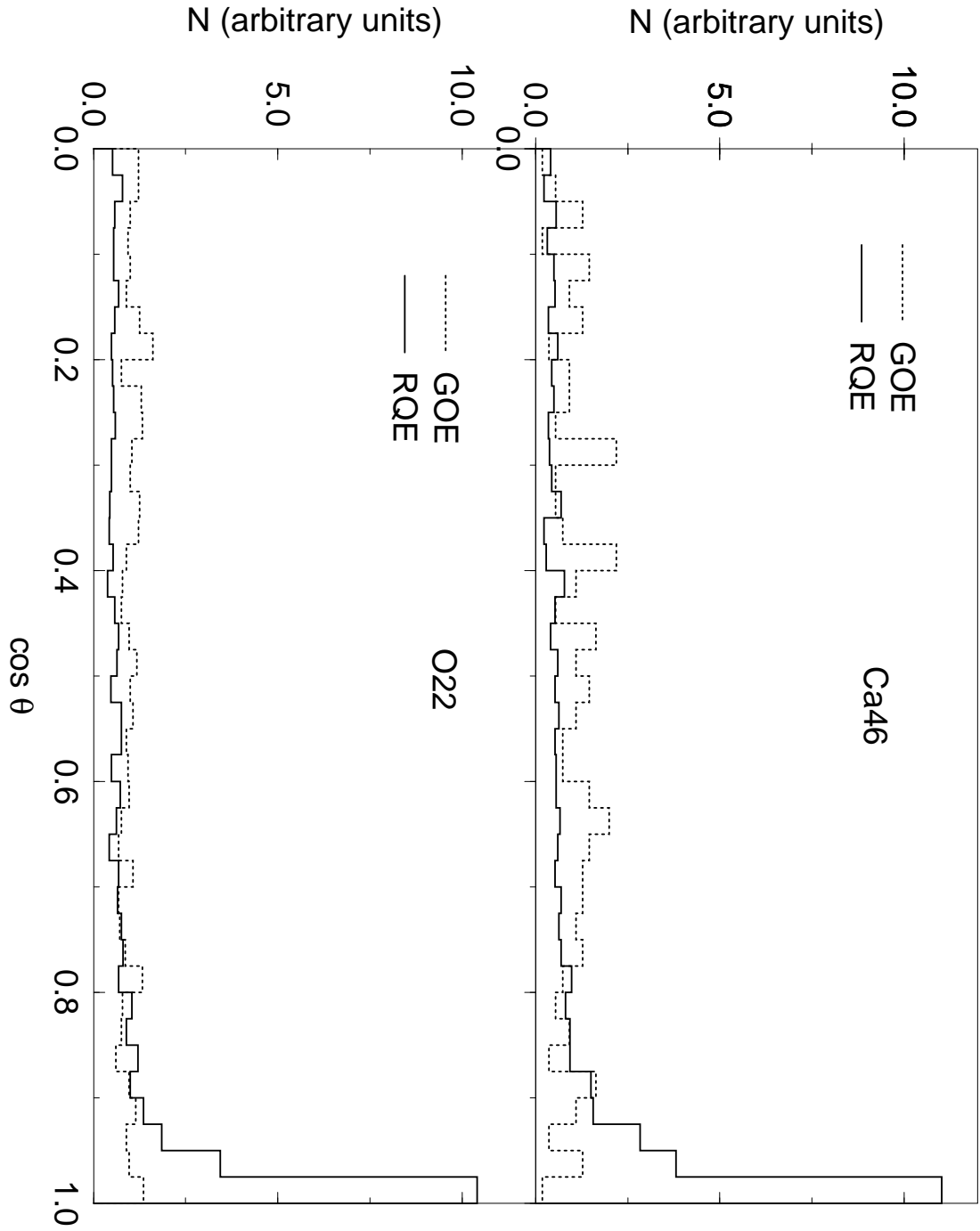,width=0.9\textwidth,angle=270}}
    \end{center}
\caption{Distribution of the correlation angle (see text for
definition) between neighboring pair-transfer amplitudes.}
\end{figure}

\begin{figure}
  \begin{center}
    \leavevmode
    \parbox{0.7\textwidth}
      {\psfig{file=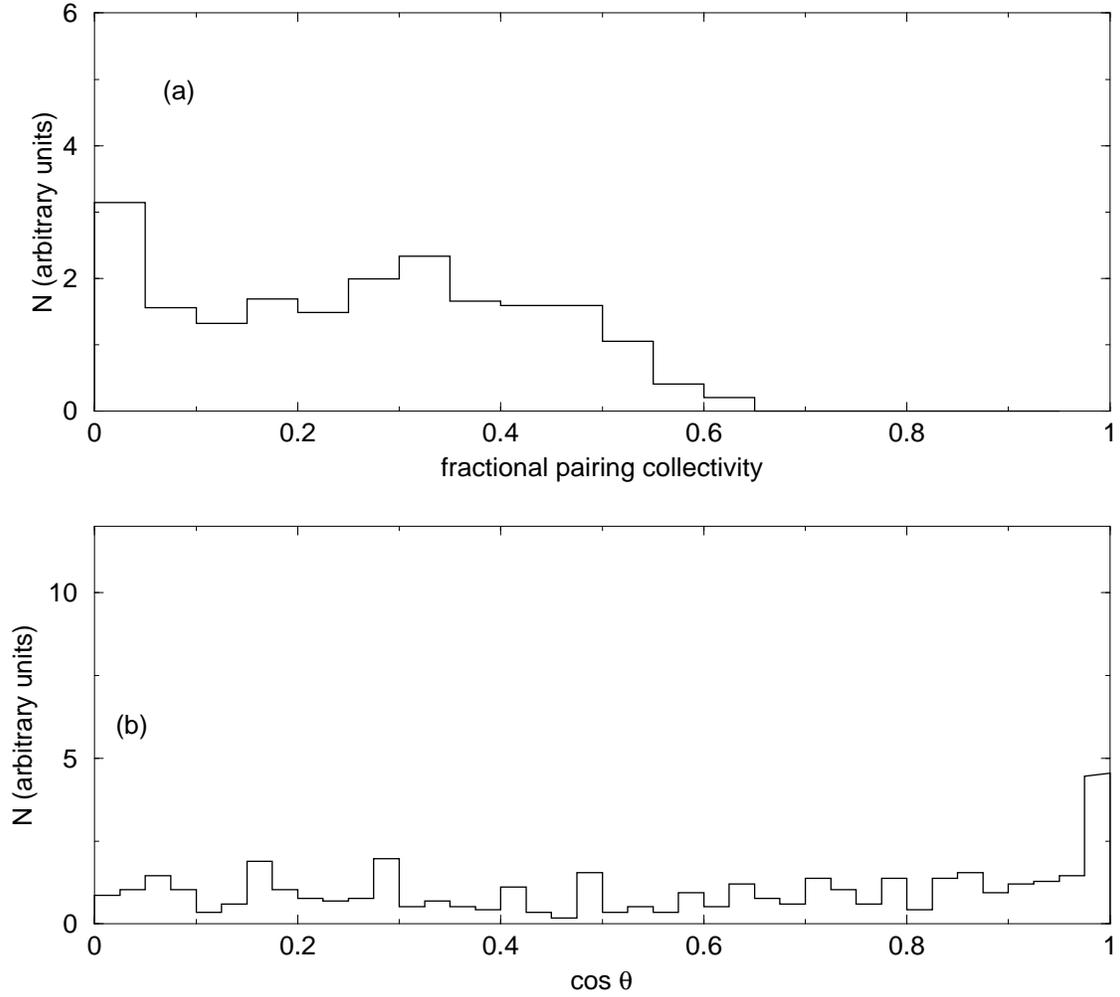,width=0.9\textwidth,angle=270}}
    \end{center}
\caption{Results for Mg isotopes. (a) Same as Fig.\ 5, for RQE.
Fractional pair-transfer collectivity for $^{26}$Mg$\rightarrow^{24}$Mg.
(b) Same as Fig.\ 6, for RQE.  Distribution of correlation angle
for $^{28}$Mg$\rightarrow^{26}$Mg and $^{26}$Mg$\rightarrow^{24}$Mg.
}
\end{figure}

\end{document}